\documentclass[twocolumn,a4paper,amsmath,amssymb]{revtex4}
\usepackage{graphicx}
\usepackage{amsmath}

\usepackage{braket}
\usepackage{xcolor}
\usepackage{natbib}
\usepackage{dsfont}

\renewcommand{\eqref}[1]{Eq.~(\ref{#1})}

\newcommand{\figref}[1]{Fig.~\ref{#1}}

\newcommand{\removedD}[1]{{\color{gray}{#1}}}
\renewcommand{\removedD}[1]{{}} 
\newcommand{\CE}[1]{{\color{black}{#1}}}
\newcommand{\Cite}[1]{\cite{#1}}

\begin{document}
\title{Characterizing Quantum Microwave Radiation \\ and its Entanglement with Superconducting Qubits using Linear Detectors}
\author{C. Eichler, D. Bozyigit and A. Wallraff}
\affiliation{Department of Physics, ETH Z\"urich, CH-8093, Z\"urich, Switzerland.}
\date{\today}
\begin{abstract}
Recent progress in the development of superconducting circuits has enabled the realization of interesting sources of nonclassical radiation at microwave frequencies. Here, we discuss field quadrature detection schemes for the experimental characterization of itinerant microwave photon fields and their entanglement correlations with stationary qubits. In particular, we present joint state tomography methods of a radiation field mode and a two-level system. Including the case of finite quantum detection efficiency, we relate measured photon field statistics to generalized quasi-probability distributions and statistical moments \CE{for one-channel and two-channel detection}. We also present maximum-likelihood methods to reconstruct density matrices from measured field quadrature histograms. \CE{Our theoretical investigations are supported by the presentation of experimental data, for which microwave quantum fields beyond the single-photon and Gaussian level have been prepared and reconstructed}.
\end{abstract}
\maketitle
\section{Introduction}
Microwave frequency quantum fields confined in cavities have been generated and characterized with remarkable control using Rydberg atoms and superconducting qubits for state preparation and readout. These experiments have illuminated fundamental principles of quantum physics, e.g.~by exploring the coherent superposition of quantum states \cite{Deleglise2008,Hofheinz2009} and their decoherence \cite{Brune1996,Deleglise2008,Wang2009a}, the entanglement between multiple modes \cite{Wang2011b} and the stabilization of Fock states using quantum feedback \cite{Sayrin2011}. More recently, progress has been made in the characterization of propagating quantized microwave fields.
They have so far been prepared in squeezed \cite{Castellanos2008,Bergeal2012} and single photon states \cite{Houck2007,Astafiev2010} and fully characterized using time-correlation measurements \cite{Bozyigit2011,Lang2011} and quantum state tomography methods \cite{Mallet2011,Eichler2011,Menzel2010,Eichler2011a,Flurin2012}. These developments have also benefited from advances in the efficient detection of microwave fields. Both quantum limited linear amplifiers \cite{Yurke2006,Yurke1987,Castellanos2008,Bergeal2010,Kinion2008,Hatridge2011} and photon counters \cite{Chen2011a, Romero2009}   significantly extend the range of potential quantum optics experiments using microwave photons interacting with superconducting qubits, nanomechanical resonators, quantum dots \cite{Frey2012,Delbecq2011}, spin ensembles \cite{Schuster2010,Kubo2010} and Rydberg atoms \cite{Hogan2012}.

The use of itinerant microwave photons in quantum optics experiments requires efficient field characterization methods. A detailed understanding of microwave and optical field detection schemes allows for adapting existing quantum optics tools to the special requirements of microwave fields. In the first part of this paper we discuss field quadrature detection schemes at microwave frequencies, their optical analogue and their use for determining the quantum state of a single mode of a radiation field. We discuss the relation between measurement results of single-channel detection schemes and quasi-probability distributions. \CE{We give new insight into the microwave state tomography problem by developing a method to reconstruct the maximally-likely Fock space density matrix directly from the measured probability distributions. Furthermore, we present state tomography experiments in which quantum states beyond the single photon level have been prepared and reconstructed. We also show that two-channel microwave detection can be interpreted as a positive $P$ function measurement~\cite{Agarwal1994} even in the presence of added classical detection noise. In the final part of the paper we develop new methods, which allow for the characterization of entanglement between a localized qubit and a radiation field mode in full joint tomography.}

\begin{figure}[b]
\centering
\includegraphics[scale=0.8]{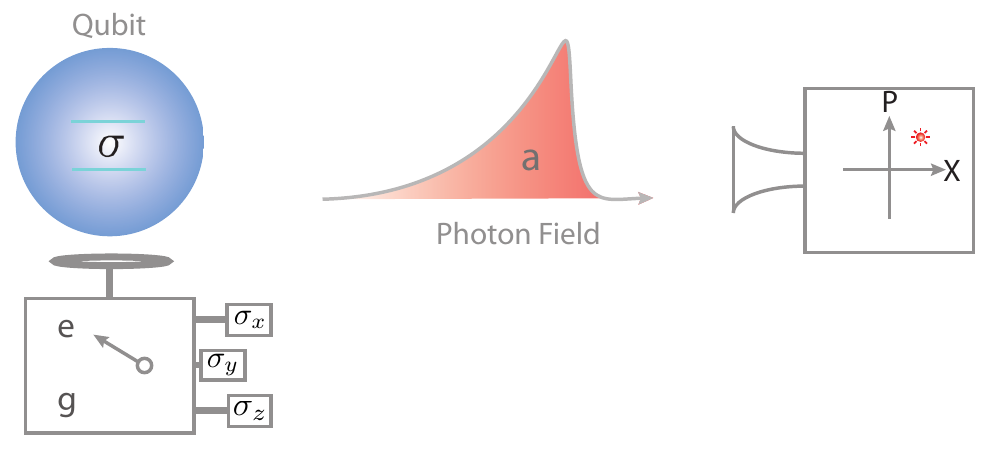}
\caption{Schematic drawing of a situation where two conjugate field quadatures $\hat{X}$ and $\hat{P}$ of a radiation field mode $a$ and an arbitrary spin component of a localized qubit ${\bf \sigma}$ are measured.
}
\label{fig:GenericSystem}
\end{figure}

The presented methods are intended for use in  state-of-the-art circuit QED experiments. However, they are also applicable in their general form to other systems described by the schematic shown in Fig.~\ref{fig:GenericSystem}. It represents the generic situation in which two canonically conjugate field quadratures $\hat{X}$ and $\hat{P}$ of a bosonic mode $a$ and an arbitrary spin component $\sigma_i$ of a localized qubit are both measured.

\section{Optical and microwave frequency field detection}
In this section we describe field quadrature detection schemes which are frequently used in the optical and in the microwave frequency range. We consider both the measurement of a single field quadrature and the simultaneous detection of two canonically conjugate quadratures. We describe the radiation field of interest as a single bosonic mode $a$ reaching the detector within a specific window in time. The single mode $a$ can be isolated from the continuum of modes by performing temporal mode matching, i.e. integrating the continuous signal over the temporal profile of the photon pulse which is to be characterized \Cite{Carmichael2008Book, Eichler2011}. We discuss ideal temporal mode matching for an exponentially decaying cavity field  in  Appendix~\ref{app:temporalmodematching}.

For a full reconstruction of the quantum state of the field both the photon number statistics and all coherences between the different contributing Fock states have to be experimentally determined. This can be achieved by measuring generalized field quadrature components $\hat{X}_{\phi}\equiv \frac{1}{2}(a e^{- i \phi}+a^\dagger e^{ i \phi})$ instead of the photon number $a^\dagger a$, which naturally allows for exploring the full phase space, i.e. the off-diagonal elements of the density matrix in the number state basis \cite{Lvovsky2009}.

\begin{figure}[b]
\centering
\includegraphics[scale=0.08]{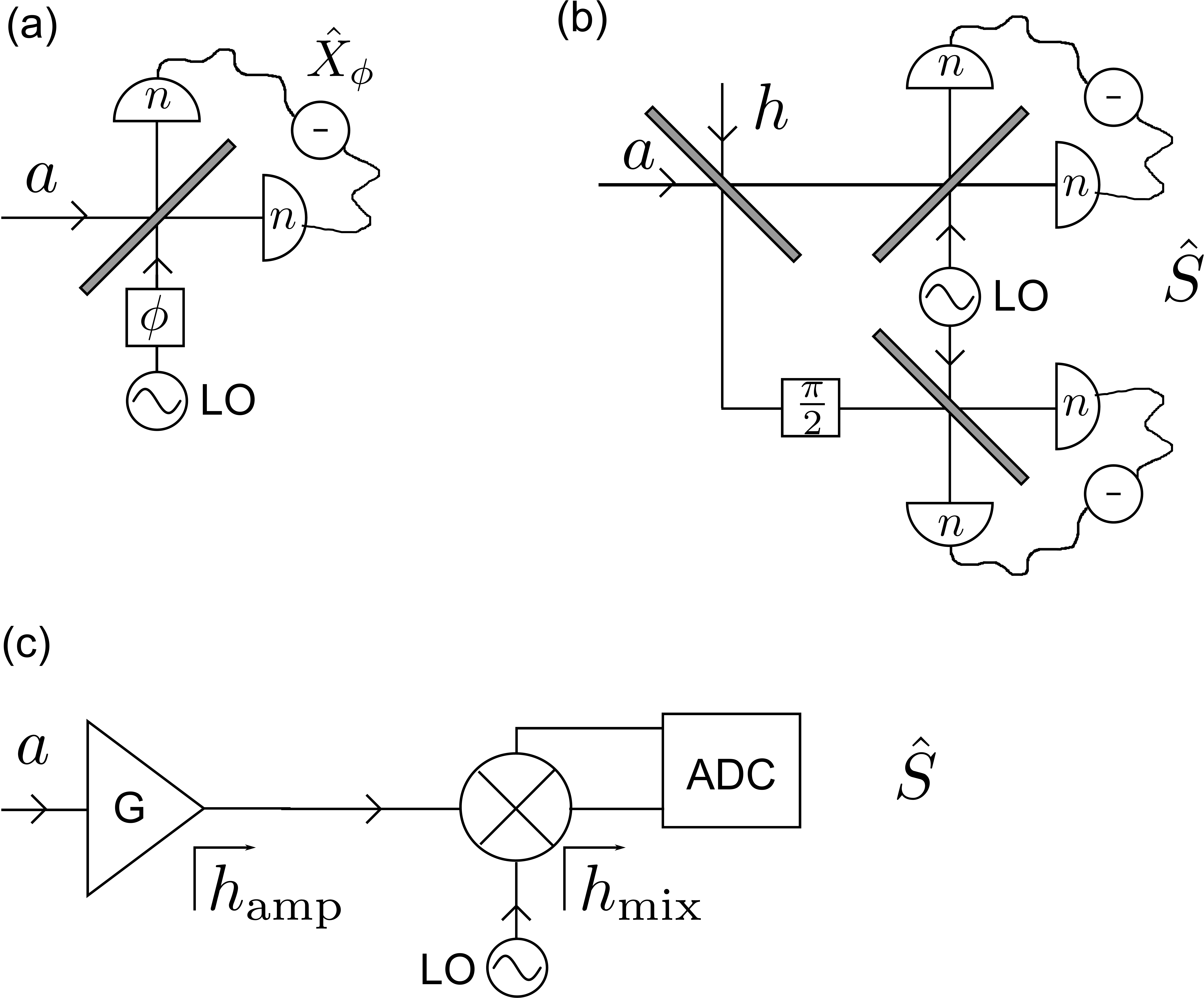}
\caption{Field quadrature detection schemes for optical and microwave fields. (a) Schematic of balanced optical homodyne detection. The signal field $a$ is combined with a coherent local oscillator (LO) field with controlled phase $\phi$ at a beamsplitter and the quadrature amplitude $\hat{X}_\phi$ is detected with photon counters (n) in the two output arms. (b) Double homodyne detection scheme. The signal field $a$ is split into two parts at a beamsplitter while introducing an additional vacuum mode $h$. Placing a homodyne detector as described in (a) at each of the two beamsplitter outputs allows for measuring two conjugate quadratures (i.e. the complex amplitude $\hat{S}$). (c) Measurement of the complex amplitude at microwave frequencies. The signal mode is amplified with a phase-insensitive linear amplifier introducing an additional noise mode $h_{\rm amp}$. At a microwave frequency mixer the amplified output is split into two parts, while adding the mode $h_{\rm mix}$, and multiplied with a coherent local oscillator field. The down-converted electrical field is sampled with analog to digital converters (ADC).}
\label{fig:quadratureDetection}
\end{figure}
In optical systems, where number statistics are naturally obtained using photon counters, such a field quadrature measurement can be realized using homodyne detection schemes \cite{Scully1997}. In this approach, the field of interest is combined on a beam-splitter with a strong coherent field of a local oscillator, such that the difference of the photocurrents at the two beamsplitter outputs are proportional to a specific field quadrature $\hat{X}_\phi$ of the input field (see Fig.~\ref{fig:quadratureDetection}(a)). The quadrature phase $\phi$ can be tuned by changing the local oscillator phase. Instead, microwave field quadratures are usually measured by down-converting the field with a local oscillator tone  using a microwave frequency mixer and sampling the electrical field directly using analog to digital converters (ADC). However, these ADCs are only sensitive enough to detect large amplitude fields which contain a macroscopic number of photons per sampling time, such that a linear amplification stage is required in the process of detection, as shown in Fig.~\ref{fig:quadratureDetection}(c). The noise added during this amplification process is typically the main limitation for the detection efficiency of microwave fields as discussed below.

Instead of measuring a single field quadrature for different phases $\phi$, two conjugate field quadratures can be simultaneously measured to get all the information required for a complete quantum state reconstruction \Cite{Arthurs1965, Braunstein1991,Caves1994, Welsch1999, Yuen1980}. One possible realization of such a measurement uses a beamsplitter and two quadrature detectors at each output \cite{Noh1991} (see Fig.~\ref{fig:quadratureDetection}(b)). The beamsplitter necessarily introduces an additional mode $h_{}$ through its open port. This mode adds [at least] the vacuum noise to the signal with which the simultaneous detection of conjugate variables preserves Heisenberg's uncertainty principle. Taking the beamsplitter transformations $a \rightarrow (a+h_{})/\sqrt{2}$ and $h_{} \rightarrow (a-h_{})/\sqrt{2}$ into account, the two detected field quadratures at the beamsplitter outputs correspond to real $\hat{X}$ and imaginary $\hat{P}$ part of the complex amplitude $a + h_{}^\dagger$. This holds for both the optical and the microwave case. However, for microwaves we still have to consider the transformation of the signal mode due to the linear amplification stage. A generic phase-insensitive linear amplifier transformation can be modeled as \Cite{Haus1962, Caves1982, Clerk2010}
\begin{equation}
a \rightarrow \sqrt{G} a + \sqrt{G-1} h_{\rm amp}^\dagger
\end{equation}
where $h_{\rm amp}$ is an additional bosonic mode accounting for the noise added by the amplifier. Again, in the ideal (i.e. quantum limited) case $h_{\rm amp}$ is in the vacuum state, and for a more realistic scenario in a thermal state. Combining the amplification transformation with the beamsplitting at the mixing stage (compare Fig.~\ref{fig:quadratureDetection}(c)) and dividing by $\sqrt{G/2}$ we find the relation \cite{daSilva2010}
\begin{equation}
\hat{S} \equiv {a} + {h}^\dagger = \hat{X} + i \hat{P}.
\label{eq:complexAmp}
\end{equation}
with the total noise mode $h = \sqrt{\frac{G-1}{G}}h_{\rm amp} + \sqrt{\frac{1}{G}}h_{\rm mix}$. Here, we have defined the complex amplitude operator $\hat{S}$ representing the two conjugate quadratures as a single complex number. In the limit of large gain $G\gg 1$ the total noise is dominated by the amplifier noise $h\approx h_{\rm amp}$ and the following noise contributions can be neglected \cite{Leonhardt1994}. Furthermore, we notice that once we amplify the field phase-insensitively at least the vacuum noise is added independently of whether we detect only one quadrature or two conjugate quadrature components. Once the signal is amplified it is thus natural to detect 2 conjugate quadratures since the signal-to-noise ratio is unaffected by the necessary splitting of the signal.

It is important to mention that there is a detection scheme using linear amplifiers which is ideally noiseless for one quadrature component. This is achieved by using a phase-sensitive amplifier instead of a phase-insensitive one which can, in the quantum limit, be modeled by the squeezing transformation \Cite{Caves1982, Loudon2000, Yurke2006}
\begin{equation}
a \rightarrow \sqrt{G} e^{- i\phi} a + \sqrt{G-1} e^{ i\phi} a^\dagger
\label{eq:phaseSensitive}
\end{equation}
with the tunable phase $\phi$. Amplifiers have recently been built which are described by this transformation and are working close to the quantum limit \Cite{Yurke1987, Castellanos2008, Hatridge2011}. The quadrature $\hat{X}_{\phi}$ is noiselessly amplified while its conjugate quadrature is deamplified. The detection scheme is thus equivalent to an optical homodyne detection \Cite{Caves1994, Lvovsky2009, Mallet2011,Filippov2011}.

We note that while for optical fields the simultaneous detection of two conjugate quadratures requires a more complicated setup than for photon number detection it is the natural measurement observable for microwave fields which we will therefore focus on in this work in the context of quantum state reconstruction.
\section{Quantum state reconstruction based on single channel field quadrature detection}\label{sec:photonTomo}
Here, we describe quantum state tomography based on the measurement of the complex field amplitude $\hat{S}$.
The goal of quantum state reconstruction is the estimation of the density matrix $\rho_{a}$ which characterizes the state of the field mode $a$. This is experimentally achieved by preparing the state many times and performing a set of measurements on these states, which contain information about the diagonal and off-diagonal elements of $\rho_a$ . Depending on the set of observables the obtained results have a direct analogy to particular representations of the density matrix. In the case of field quadrature detection these representations are phase space distributions such as the Husimi-$Q$ function or the Wigner function \cite{Cahill1969a,Carmichael1999Book,Haroche2006}. in the following we discuss their relation to statistical moments and the Fock basis representation of the density matrix.
\subsection{Phase space distributions}
Due to the non-orthogonality of coherent states $\langle \alpha|\beta \rangle = e^{-\frac{1}{2}{|\alpha|^2}-\frac{1}{2}{|\beta|^2}}e^{\alpha^* \beta}$ an arbitrary density  matrix $\rho_{a}$ can be expanded as a linear combination of projectors $|\alpha\rangle\langle\alpha|$ onto coherent states
\begin{equation}
\rho_{a}=\int_{\alpha} P_{a}(\alpha)|\alpha\rangle\langle\alpha|.
\label{eq:PFuncDef}
\end{equation}
Here we have defined $\int_{\alpha} \equiv \int_{\mathds{C}} \text{d}^2 \alpha$ for integrals over the complex plane and $P_{a}(\alpha)$ as the Glauber-Sudarshan $P$ function \Cite{Glauber1963c, Carmichael1999Book}, which uniquely represents the density matrix as a distribution in phase space. $P_{a}(\alpha)$  is always real-valued but can be negative and can contain  singularities proportional to derivatives of the Dirac $\delta$ distribution to all orders \Cite{Carmichael1999Book}. As can be seen from its definition Eq.~(\ref{eq:PFuncDef}) the $P$ function reduces to a two-dimensional Dirac distribution $P_{a}(\alpha) = \delta^{(2)}(\alpha-\beta)$ for coherent states $|\beta\rangle$. Coherent states thus appear as single points in phase space with no statistical spread similar to their classical counterparts. For this reason and due to its possible negative values the $P$ function does not directly describe the statistics of measurements. However, it is very useful since its statistical moments directly correspond to the normally ordered moments
\begin{eqnarray}
\langle (a^\dagger)^m a^n\rangle &=& \int_\alpha \,\, (\alpha^*)^m \alpha^n \,\,P_a(\alpha)\,\,
\end{eqnarray}
of the field operator $a$ and because of its analogy to probability distributions of classical fields. A second distribution, which is of particular relevance for the following discussion, is the Husimi-$Q$ function \begin{equation}
Q_{a}(\alpha)=\frac{1}{\pi}\langle\alpha|\rho_a|\alpha\rangle,
\label{eq:QFuncDef}
\end{equation}
since it generates the anti-normally ordered moments
\begin{eqnarray}
\langle  a^n (a^\dagger)^m\rangle&=& \int_\alpha \,\, (\alpha^*)^m \alpha^n \,\,Q_a(\alpha)\,\,.
\end{eqnarray}
Substituting Eq.~(\ref{eq:PFuncDef}) into the definition of the $Q$ function, we note that it is related to the $P$ function by a Gaussian convolution. For coherent states $Q_{a}(\alpha)$ becomes a two-dimensional Gaussian distribution with variance 1 centered around the coherent state amplitude. Half of these fluctuations describe the intrinsic vacuum fluctuations of the quantum field, the other half describe the minimal added uncertainty when directly measuring a $Q$ function, which requires the simultaneous detection of two conjugate field quadratures.

Both distributions are special cases of the \emph{s-parametrized} quasi-probability  distribution (QPD) $W_{a}(\alpha,s)$
\begin{align}
Q_{a}(\alpha) &= W_{a}(\alpha,-1)\\
P_{a}(\alpha) &= W_{a}(\alpha,1).
\end{align}
which has been introduced by Cahill and Glauber \Cite{Cahill1969a} as a generalized phase space representation of the density matrix where the parameter $s\in (-\infty,+1]$. For different values of $s$ the QPDs are related to each other by a Gaussian convolution \Cite{Cahill1969a}
\begin{align}\label{eq:QPDDef}
 W_{a}(\alpha,s) = \frac{2 \pi^{-1}}{t-s} \int_\beta \exp \left( -\frac{2|\alpha-\beta|^2}{t-s} \right) W_{a}(\beta, t)
\end{align}
where $t>s$.

An intuitive interpretation of the parameter $s$ relates to the amount of fluctuations which are contained in the distribution in units of half photons. For $s=0$ we obtain the Wigner function $W_{a}(\alpha)\equiv W_{a}(\alpha,0)$  which include the intrinsic vacuum fluctuations but no additional noise due to measurement. In the case $s=1$ we identify the $P$ function where even the vacuum fluctuations are not represented. On the other hand, for $s=-1$ the QPD corresponds to the $Q$ function where both the vacuum fluctuations and the minimal added  detection noise are embedded. As discussed below,  additional classical detection noise leads to $s<-1$ when identifying measured distributions with a generalized QPD.
\subsection{Measurement of phase space distributions}\label{sec:PS}
In order to understand the relation between generalized QPDs and measured distributions in different experimental situations we assume that the complex amplitude $\hat{S} = a + h^\dagger$ as introduced above is the measured observable and that the results $S$ of repeated measurements are stored in a 2-dimensional histogram $D^{[\rho_a]}(S)$ where the two axes represent the real and imaginary part of $S$.
From this measured distributions all statistical moments can be numerically evaluated as
\begin{eqnarray}
\langle(\hat{S}^\dagger)^n \hat{S}^m  \rangle_{\rho_a} &=& \int_S \,\, (S^*)^n S^m \,\,D^{[\rho_a]}(S)\,\,.
\label{eq:moments}
\end{eqnarray}
If the noise added by the detection chain is independent of the signal generated by the photon source the signal mode $a$ and the noise mode $h$ are uncorrelated $\rho = \rho_{a}\otimes \rho_{h}$. Under this assumption the moments of the measured distribution can be decomposed into products of signal and noise moments
\begin{eqnarray}
\langle(\hat{S}^\dagger)^n \hat{S}^m \rangle_{\rho_a} &=&
\sum_{i,j=0}^{m,n} \binom{n}{j}\binom{m}{i}
\langle(h^\dagger)^i h^j  \rangle
\langle a^{m-i} (a^\dagger)^{n-j}\rangle,
\nonumber
\\
\label{eq:ON}
\end{eqnarray}
Here, we have chosen an operator ordering where the signal moments $\langle a^m (a^\dagger)^n \rangle$ appear anti-normally ordered and the noise moments $\langle(h^\dagger)^m h^n  \rangle$ normally ordered. Note that since $\hat{S}$ is a normal operator $[\hat{S},\hat{S}^\dagger]=0$ one can express Eq.~(\ref{eq:ON}) also with opposite ordering as shown in Eq.~(\ref{eq:ON2}) later in the text.

The probability distribution for the sum of two independent random variables $a+h^\dagger$ is identical to the convolution of the individual distributions for $a$ and $h^\dagger$. As a result, one possible representation of the probability distribution $D^{[\rho_a]}(S)$  is given by the convolution \Cite{Kim1997}
\begin{eqnarray}
D^{[\rho_a]}(S) = \int_\alpha \, P_h(S^* - \alpha^*) Q_a(\alpha).
\label{eq:convolution}
\end{eqnarray}
In the following we discuss special cases of Eq.~(\ref{eq:convolution}). At optical frequencies the measurement of $\hat{S}$ can be realized using a double homodyne or heterodyne detection and the noise mode $h$ is nearly in the vacuum state for which $P_{h}(\beta)=\delta^{(2)}(\beta)$ resulting in
\begin{equation}
D^{[\rho_a]}(S) = Q_a(S).
\end{equation}
In contrast, for microwave fields the noise mode $h$ is often in a thermal state with mean photon number $N_{ 0}$ ranging typically from 0.5 to 10 if parametric or SQUID amplifiers are used \Cite{Mallet2011, Vijay2011, Kinion2008} or between 30 an 200 if the first amplification is performed by a transistor based amplifier \Cite{Bozyigit2011, Eichler2011, Eichler2011a, Lang2011}. In this case $P_{h}(\alpha)=e^{- |\alpha|^2 /N_{0}}/\pi N_{0}$ acts as a Gaussian filter and by comparing with  Eq.~(\ref{eq:QPDDef}) we obtain the broadened QPD
\begin{equation}
D^{[\rho_a]}(S) = W_a(S,-1-2N_{0}).
\end{equation}
Note that finite thermal noise in $h$ can be equivalently interpreted as optical homodyne detection with finite detection efficiency $\eta$ for which the measured distribution is given by $D^{[\rho_a]}(S) = W_a(S,1-2\eta^{-1})$ \Cite{Leonhardt1993}. Added noise can thus be understood as a reduced detection efficiency $\eta = 1/(1+N_{0})$.

We conclude that under the experimentally verified assumption \cite{Eichler2011,Menzel2010} of $h$ being in a thermal state not correlated with $a$ the measured distribution of $\hat{S}$ is a direct measurement of the generalized QPD and therefore contains all information required to reconstruct the density matrix $\rho_a$ of the state of interest or to test its nonclassical properties \cite{Vogel2000,Kiesel2011}. In contrast to other reconstruction schemes only a single observable $\hat{S}$ needs to be measured.

However, in many experiments the mean photon number of the noise field is larger than the mean photon number of the signal field $N_{0} > \langle a^\dagger a\rangle$ and consequently the features of measured probability distributions are on first sight dominated by the noise distribution. Therefore the goal is to systematically extract the information contained about mode $a$ in the measured QPD and represent it in a form, which allows for a direct estimation of the properties of the state, such as the fidelity with respect to an expected density matrix.
\subsection{Determination of normally ordered moments}
One way of quantifying the properties of a quantum state is to analyze the statistical moments $\langle (a^\dagger)^n a^m \rangle$ of the field operator \cite{Buzek1996,Menzel2010}, since quantities such as the mean amplitude, the mean photon number and the variance in the photon number can be extracted immediately. In this section we discuss the approach developed in Ref.~\Cite{Eichler2011} to extract these moments from the measured distributions in the presence of significant amplifier noise $N_{0}$. The basic idea is to deconvolve the QPDs for the field operators $a$ and $h$ order by order.

Rewriting Eq.~(\ref{eq:ON}) with a different choice of operator ordering
\begin{eqnarray}
\langle(\hat{S}^\dagger)^n \hat{S}^m \rangle_{\rho_a} &=&
\sum_{i,j=0}^{n,m}\binom{m}{j}\binom{n}{i}
\langle(a^\dagger)^i a^j  \rangle
\langle h^{n-i} (h^\dagger)^{m-j}\rangle,
\nonumber
\\
\label{eq:ON2}
\end{eqnarray}
we find that once the anti-normally ordered moments of the noise mode $\langle h^{n} (h^\dagger)^{m}\rangle$ are known, the set of linear equations can be solved for  $\langle (a^\dagger)^n a^m \rangle$. From Eq.~(\ref{eq:moments}) we note that a reference measurement $D^{[|0\rangle\langle0|]}(S)$, for which $a$ is prepared in the vacuum, gives direct access to the moments $\langle h^{n} (h^\dagger)^{m}\rangle$, since all normally ordered moments in $a$ with $n,m \neq 0$ are then $\langle (a^\dagger)^n a^m \rangle =0$ and Eq.~(\ref{eq:ON2}) reduces to
\begin{equation}
\langle(\hat{S}^\dagger)^n \hat{S}^m  \rangle_{|0\rangle \langle 0|} = \langle h^{n} (h^\dagger)^{m}\rangle \,\,.
\label{eq:OFF}
\end{equation}
In cryogenic setups such a reference measurement with $a$ in the vacuum can typically be performed by  cooling the source of radiation into the ground state or very close to it \Cite{Fink2010}.
The identity in Eq.~(\ref{eq:OFF}) can be understood as follows: The situation with $a$ in the vacuum state corresponds to an ideal $Q$ function measurement for the noise mode $h$ and the moments generated by this distribution are exactly the anti-normally ordered ones appearing in Eq.~(\ref{eq:OFF}). We finally invert Eq.~(\ref{eq:ON2}) to extract the desired moments $\langle (a^\dagger)^n a^m \rangle$ of the mode to be characterized.

In principle, the moments of the measured histograms can be evaluated to arbitrary order.  However, there are limitations in the accuracy with which the moments $\langle (a^\dagger)^n a^m \rangle$ can be determined depending on the integration time and the detection efficiency. As investigated theoretically in Ref.~\Cite{daSilva2010}, the statistical error of the moments increases with increasing order. The result shows that
the number of measurements which are necessary to extract a moment of order $M$ with a given precision scales with $(1+N_{0})^{M}$. The measurement time necessary to determine higher order moments with a fixed precision thus scales exponentially with increasing order.

The state of a single mode of the radiation field has an infinite number of degrees of freedom, i.e. an infinite dimensional Hilbert space. This makes it in principle impossible to exactly reconstruct a state, because an infinite amount of information is to be acquired. However, the measurement of a finite set of moments often allows for a controlled reduction of the relevant state space \Cite{Buzek1996}.

\subsection{Special classes of states and the Fock space density matrix}
One class of states which is characterized by a finite set of moments comprises  coherent, thermal and squeezed (i.e. Gaussian) states, for which the statistical moments up to second order
\begin{align}
\left\{ \left\langle a \right\rangle,\left\langle a^\dag a \right\rangle, \left\langle a^2 \right\rangle \right\}.
\end{align}
determine all higher order moments.  In order to analyze how close the reconstructed state really is to a Gaussian, one has to measure the third order cumulants and evaluate their deviations from zero.

A second class of states which can be reconstructed using a finite set of measured moments includes those with finite photon number contributions satisfying $\langle n | \rho_a |m \rangle = 0$ for $m,n \geq N$ in the Fock basis $\{|n\rangle \}$. For these states the normally ordered moments
\begin{align}
\left\langle (a^\dag)^n a^m \right\rangle = 0 \quad m \text{ or } n \geq N
\end{align}
vanish and the state is completely determined by the finite set of moments
\begin{align}
\left\{ \left\langle (a^\dag)^n a^m \right\rangle \right\} \quad m \text{ and } n \leq N.
\end{align}
It is important to note that it necessarily follows from $\langle (a^\dag)^N a^N\rangle = 0$  that there are no Fock states $|n\rangle$ with $n\geq N$ contributing to the density matrix. If $\langle (a^\dag)^N a^N\rangle < \epsilon $ can be verified experimentally one knows an upper bound
\begin{equation}
\epsilon > \langle (a^\dag)^N a^N \rangle = \sum_{n\geq N} \langle n|\rho_{a}|n\rangle \frac{n!}{(n-N)!} \geq \sum_{n\geq N}\langle n|\rho_{a}|n\rangle
\label{eq:inEq}
\end{equation}
for the sum of higher order Fock state populations. The approximation made when truncating the Hilbert space is thus well-controlled.

If such a truncation is possible the moments can be mapped to a density matrix in Fock representation by evaluating the sum \Cite{Herzog1996}
\begin{eqnarray}
\langle m|\rho_{a}|n\rangle &=& \frac{1}{\sqrt{n!m!}}\sum_{l=0}^{\infty} \frac{(-1)^{l}}{l!}\langle(a^\dagger)^{n+l} a^{m+l}\rangle
\nonumber
\\
&\equiv& \mathcal{M}(\langle(a^\dagger)^n a^m\rangle)
\label{eq:momsToDensMat}
\end{eqnarray}
up to terms of order $2N$.

The described procedure is very efficient since the evaluation of moments from the measured distributions as well as finding the solution of Eq.~(\ref{eq:ON2}) requires only small computational effort. Furthermore the moment representation provides a very intuitive picture to extract fundamental properties of the quantum state.
%
\section{Maximum-likelihood state estimation based on generalized complex amplitude detection}
Due to the unavoidable statistical imprecision in expectation values extracted from a finite number of measurements, a direct mapping from the measurement data to the desired state representation does not in general result in a completely positive density matrix. Maximum-likelihood state estimation  aims to correct for that. In this section we discuss two different maximum-likelihood procedures applicable to complex amplitude detection schemes as relevant for the circuit QED experiments under consideration. The first method is based on the experimentally determined finite set of moments $\langle  (a^\dagger)^{n}a^{m}\rangle$ together with their respective standard deviations $\delta_{n,m}$. The second one estimates the density matrix directly from the measured probability distributions.
\subsection{Maximum-likelihood procedure based on measured moments}

In order to find the most likely density matrix given a set of measured moments and their respective standard deviations $\delta_{n,m}$ we maximize the log-likelihood function \cite{Chow2012}
\begin{equation}
\mathcal{L_{\rm Log}} = - \sum_{n,m}\frac{1}{\delta_{n,m}^2}|\langle  (a^\dagger)^{n}a^{m}\rangle - {\rm Tr}[\rho_a (a^\dagger)^n a^m]|^2
\label{eq:LogL}
\end{equation}
with respect to the elements of the density matrix $\rho_a$. The properties $\rho_a \geq 0$ and ${\rm Tr}\rho_a=1$ of the density  matrix are included as constraints in the maximization of  Eq.~(\ref{eq:LogL}). The standard deviations $\delta_{n,m}$ appear in the denominator of each term, such that moments which are determined with low accuracy contribute to the log-likelihood function with less weight.

This maximization problem can be formulated as a semi-definite program, for which efficient numerical solutions exist \cite{Vandenberghe1996,Chow2012}.  Note that this maximum likelihood scheme is particularly efficient for states which contain only few photons since in this case only a finite set of moments is non-zero.

\begin{figure*}
\includegraphics[scale=1]{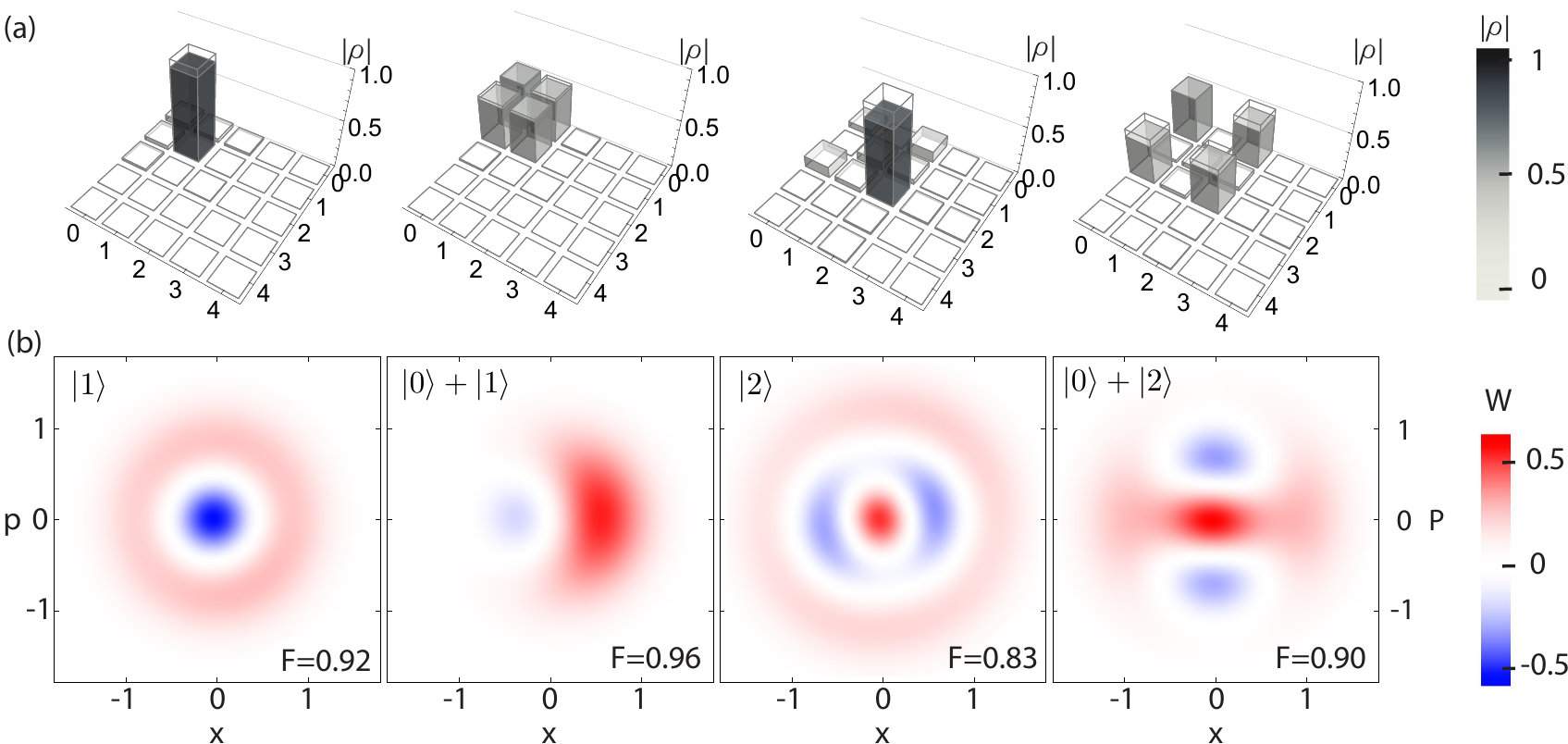}
\caption{ (a) Absolute value of the experimentally reconstructed density matrices (grayscale) in comparison with the ideal ones (wireframes) for the four indicated quantum states. (b) Measured density matrices  transformed into their corresponding Wigner functions $W(x,p)$. The fidelities between ideal states $|\psi\rangle$ and measured density matrices are evaluated as  $F=\langle\psi|\rho|\psi\rangle$.
}
\label{fig:densMatFromMoms}
\end{figure*}

\CE{ We have tested the described maximum-likelihood procedure based on experimental data sets obtained in a circuit QED experiment. In addition to the generation of single photon states \cite{Eichler2011} we have prepared two photon Fock states and their coherent superposition with the vacuum. Note that for the reconstruction of 2 photon states it is necessary to measure photon correlations including moments up to sixth order. The accurate measurement of $\langle (a^\dagger)^3 a^3\rangle$ -- compared to previous measurements in which $\langle (a^\dagger)^2 a^2\rangle$ had been the highest order measured moment \cite{Bozyigit2011,Eichler2011,Lang2011} -- was enabled by a Josephson parametric amplifier used as the first amplifier in the detection chain \cite{Castellanos2008,Eichler2012a}.

Based on the measured moments and their respective standard deviations up to order $n+m=8$ we reconstruct each density matrix by maximizing Eq.~(\ref{eq:LogL}). In order to demonstrate that higher order photon number populations are not relevant for the description of the state if one of the diagonal moments $(a^\dagger)^N a^N\rangle$ is measured to be close to zero (compare Eq.(\ref{eq:inEq})), we have chosen a Hilbert space with up to four photon Fock states. The results (see Fig.~\ref{fig:densMatFromMoms}a) show that only the zero, one and two photon Fock states contribute to the reconstructed density matrices while the higher Fock states stay  unpopulated. A compromise between the size of the Hilbert space and the likelihood of the reconstructed state may be found by applying the Akaike or Bayesian information criterion \Cite{Guta2012} to reduce the complexity of the model used for reconstructing the state.  In order to illustrate the quantum character of the reconstructed states we have transformed the density matrices into their corresponding Wigner functions \cite{Eichler2011}, which show  negative values in all four cases (see Fig.~\ref{fig:densMatFromMoms}b).

We estimate the statistical error in the fidelities of the reconstructed density matrices by repeating the likelihood maximization for resampled sets of moments \cite{Chow2012,Rehacek2008}. The resulting standard deviations of the resampled fidelities are below $2\%$ for all  reconstructed states. The small statistical errors are due to the high overall microwave detection efficiency of $\eta = 0.19$ of our setup in combination with the large number of measurements exceeding $10^8$ for all the shown density matrices. Since high repetition rates of up to 10 MHz \cite{Bozyigit2011} are possible for circuit QED experiments we believe that the maximum likelihood approach is well suited in this context. However, in experiments for which only a small number of samples is available alternative methods such as the Bayesian approach \cite{Audenaert2009,Dobek2011} may be advantageous compared to maximum likelihood procedures.}
\subsection{Iterative maximum-likelihood procedure based on measured histograms}
In addition to the moments based maximum likelihood scheme we formulate an iterative procedure which estimates the density matrix directly from the measured histograms. This reconstruction method is useful for photon states which contain a large number of contributing Fock states and consequently a large number of non-vanishing moments. In addition to this practical relevance  it gives insight into the interpretation of the measured probability distribution.

The measurement of a quantum observable can be described by a set of positive operator valued measures (POVM) ${\hat{\Pi}}_{j}$ \cite{Nielsen2000}, which have the property that the probability $p_{j}$ for getting the respective measurement result is given by $p_{j} = {\rm Tr}[\rho \hat{\Pi}_{j}]$. The operators ${\hat{\Pi}}_{j}$ need to be positive and hermitian but not necessarily projectors. In the ideal case they form a decomposition of the Hilbert space $\sum_{j} {\hat{\Pi}}_{j} = \mathds{1}$.
Preparing and measuring a system in state $\rho$ repeatedly will return each of the possible results $f_{j}$ times. The most likely state $\rho_{\rm ML}$ given this set of data is the one which maximizes the likelihood function
\begin{equation}
\mathcal{L} = \prod_{j} {\rm Tr}[\rho \hat{\Pi}_{j}]^{f_j}.
\end{equation}
Note that in order to find a unique global maximum of $\mathcal{L}$, it is a necessary condition that an arbitrary density matrix can be constructed as a linear combination of $\hat{\Pi}_{j}$. As a counterexample, if the POVM are given by a complete orthogonal set of projectors ${\hat{\Pi}}_{j} =|j \rangle \langle j|$ the ML function $\mathcal{L}$  is independent of the off-diagonal elements of $\rho$ expressed in the $|j \rangle$ basis. The maximization of $\mathcal{L}$ can thus only identify the most likely diagonal density matrix elements $\langle j|\rho|j \rangle$.

It is computationally demanding to directly determine  $\rho_{\rm ML}$ for high-dimensional Hilbert spaces. However,  the density matrix $\rho_{\rm ML}$ can be found using iterative methods \cite{Hradil2004,Lvovsky2004,Rehacek2007,Lvovsky2009}. In order to formulate the ML iteration procedure we define the operator
\begin{equation}
\hat{R}(\rho) = \sum_{j}\frac{f_j}{{\rm Tr}[\rho \hat{\Pi}_{j}]}\hat{\Pi}_{j}.
\end{equation}
The iterative method for updating the density matrix \cite{Hradil2004,Hradil1997}
\begin{equation}
\rho_{k+1} = \mathcal{N} \hat{R}(\rho_{k}) \rho_{k} \hat{R} (\rho_{k})
\label{eq:iteration}
\end{equation}
with renormalization constant $\mathcal{N}$, has shown good convergence towards $\rho_{\rm ML}$ \cite{Lvovsky2009}.
As an initial condition for the iteration procedure one either chooses the maximally mixed state $\rho_{0}=\mathds{1}/d$, where $d$ is the dimension of the reconstructed Hilbert space, or constructs a more realistic initial condition by taking into account the measured moments.

In a practical implementation where the phase space is discretized and the Hilbert space is truncated to finite dimensions we might also be faced with the situation that the POVM operators do not sum to the identity operator $\sum_{j} \hat{\Pi}_j = \hat{G} \neq \mathds{1}$. In this situation the iteration procedure can be modified as
\begin{equation}
\rho_{k+1} = \mathcal{N} \hat{G}^{-1} \hat{R}(\rho_{k}) \rho_{k} \hat{R} (\rho_{k}) \hat{G}^{-1}
\label{eq:iteration}
\end{equation}
to guarantee convergence towards the most likely density matrix \cite{Hradil2006,Mogilevtsev2007}.
\subsubsection{Iterative method for ideal complex amplitude detection}
The method described above has been adapted to optical homodyne detection by Lvovsky \cite{Lvovsky2004} in 2004 and is frequently used in experiments based on optical homodyne tomography \cite{Babichev2004,Tipsmark2011,Usuga2010}. Here we adapt the method to measurements of the complex amplitude operator $\hat{S}$. We start with the case of ideal detection, i.e. for the noise mode $h$ being in the vacuum state.

As discussed in Section \ref{sec:PS} the measured probability distribution in this case is the $Q$ function $D^{[\rho_a]}(S) = Q_a(S)$. The underlying set of POVMs $\hat{\Pi}_{S}$ is thus defined by the condition
\begin{equation}
Q_a(S) \doteq {\rm Tr}[\rho_a \hat{\Pi}_{S}].
\end{equation}
Since the $Q$ function can be written as the expectation value $Q_a(\alpha) = \frac{1}{\pi}\langle \alpha|\rho_a|\alpha \rangle$ with respect to coherent states $|\alpha\rangle$ we identify the well-known result \cite{Helstrom1976Book}
\begin{equation}
\hat{\Pi}_{S=\alpha} \equiv \hat{\Pi}_{\alpha} = \frac{1}{\pi}|\alpha\rangle\langle\alpha|.
\end{equation}
Here and in the following we have labeled the possible measurement results of $\hat{S}$ by $\alpha$ to emphasize their relation to coherent states.

The coherent state projectors $\hat{\Pi}_{\alpha}$ have both the desired properties: They sum up to the identity operator $\int_\alpha \Pi_{\alpha} =\mathds{1}$ and they allow for the construction of an arbitrary density matrix as a linear combination of projectors
\begin{equation}
\rho_a = \pi \int_{\alpha} P_{a}(\alpha) \hat{\Pi}_{\alpha},
\end{equation}
compare with Eq.~(\ref{eq:PFuncDef}). Based on this knowledge we can directly apply the iteration procedure Eq.~(\ref{eq:iteration}).

Full state tomography thus requires the measurement of only a single observable $\hat{S}$ which ideally projects onto coherent states. Due to the properties of coherent states all information about the phase of the field necessary to reconstruct the off-diagonal density matrix elements is contained in this measurement. This is one of the reasons why the discussed detection scheme has great potential in microwave photon field tomography -- especially since the advent of nearly quantum-limited amplifiers \cite{Bergeal2010,Castellanos2008,Eichler2011a,Hatridge2011}.
\subsubsection{Iterative method for generalized complex amplitude detection}
Due to noise added by amplifiers as well as finite radiation losses in waveguides and microwave components the mode $h$ is typically not described by the vacuum but a thermal state with mean photon number $N_{0}$. In the following we show how to reconstruct the density matrix $\rho_a$ in this situation. We keep the discussion as general as possible and allow for mode $h$ being in an arbitrary state described by $\rho_h$ which can be specified experimentally using a reference measurement. \CE{The following procedure has to the best of our knowledge not yet been discussed in literature.}

Preparing the signal mode $a$ in the vacuum state we can measure the $Q$ function of mode $h$ since $D^{[|0\rangle\langle0|]}(\alpha) = Q_{h}(\alpha^*)$. Applying the iterative maximum likelihood scheme for ideal detection we reconstruct the most likely state for the  noise mode $\rho_{h}$. To account for this noise state in the reconstruction of $\rho_a$ we identify the modified POVM operators $\hat{\Pi}_{\alpha}^{[\rho_h]}$, which describe the measurement process under the condition that the detection system is in state $\rho_h$. The result, which can be shown by verifying the identity
\begin{equation}
{\rm Tr}[\rho_a \hat{\Pi}_{\alpha}^{[\rho_h]}] \doteq D^{[\rho_a]}(\alpha) \overset{\text{Eq.~(\ref{eq:convolution})}}{=} \int_\beta  P_h(\alpha^* - \beta^*)  Q_a(\beta),
\end{equation}
between POVMs and the expected measured distribution, is
\begin{equation}
\hat{\Pi}_{\alpha}^{[\rho_h]} = \frac{1}{\pi} T_{h}(\alpha)\tilde{\rho}_{h}T_{h}^\dagger(\alpha).
\label{eq:POVM}
\end{equation}
Here we have defined the displacement operator $T_{h}(\alpha) \equiv e^{\alpha h^\dagger-\alpha^* h}$ and $\tilde{\rho}_h$ as the most likely density matrix  with respect to the reflected histogram $Q_{h}(-\alpha^*)$. Note that since displaced Fock states are orthonormal and complete \cite{Wuensche1991}
\begin{equation}
\int_{\alpha} T(\alpha) |m\rangle \langle n| T^\dagger(\alpha)  = \mathds{1} \delta_{n,m},
\end{equation}
the relation $\int_\alpha \hat{\Pi}_{\alpha}^{[\rho_h]}  = \mathds{1}$ holds for any valid detector state $\rho_h$. This leads to the remarkable result that the reconstruction method is unbiased for arbitrary detector states. The only two requirements for the method to apply are that the signal and noise modes are uncorrelated $\rho=\rho_a \otimes \rho_h$ and that $a$ can be cooled into the vacuum state or any other known state. Both of these conditions can be realized experimentally to good approximation as discussed before.  In order to estimate the density matrix $\rho_{a}$ we can again apply the iterative method using $\hat{\Pi}_{\alpha}^{[\rho_h]}$ as a set of POVMs.

\begin{figure}[b]
\centering
\includegraphics[scale=1]{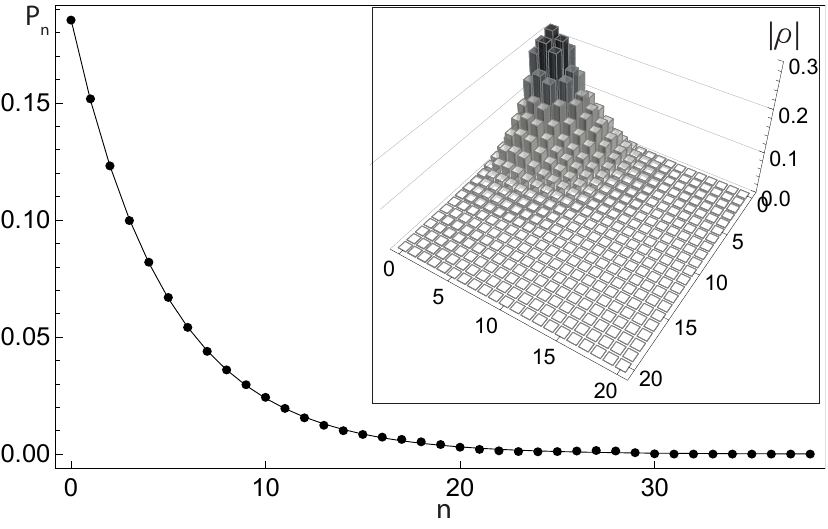}
\caption{Diagonal matrix elements $p_n =\langle n|\tilde{\rho}_h|n\rangle$ of the detector state (dots) obtained with the iterative maximum likelihood method from experimental data. The photon number distribution is well described by a thermal distribution (solid line) with mean photon number $N_0\approx4.4$. The inset shows the reconstructed density matrix with fidelity $F=95\%$ of a coherent state with $\alpha \approx 1.7$.
}
\label{fig:DensMatCoherent}
\end{figure}
\CE{We have applied the iterative maximum likelihood scheme to the same data sets as presented in Fig.\ref{fig:densMatFromMoms} and found quantitative agreement between the two methods to about  $1\%$.  As described in the following, we have  tested the iterative method also for a coherent state $|\alpha\rangle$ with mean amplitude $ \alpha\approx1.7$, for which we  expect higher photon number states to be occupied.
We first apply the iterative procedure to the reference histogram which characterizes the detector state $\tilde{\rho}_h$. Its diagonal elements $p_n =\langle n|\tilde{\rho}_h|n\rangle$ are shown in Fig.~\ref{fig:DensMatCoherent} as dots which are very well described by a thermal distribution (solid line) with mean photon number $N_0\approx4.4$. The off-diagonal elements (not shown) are all smaller than $\epsilon=0.004$.} Therefore, the detection noise is very well approximated by thermal noise. Taking into account the estimated detector state $\tilde{\rho}_h$ we construct $\hat{\Pi}_{\alpha}^{[\rho_h]}$ and iterate the maximum likelihood procedure for the coherent state histogram. The resulting estimated density matrix $\rho_a$ is shown in the inset of Fig.~\ref{fig:DensMatCoherent} and has a fidelity of $F=95\%$ compared to an ideal coherent state.

Note that in order to reconstruct and express the density matrix of the detector state $\tilde{\rho}_h$ with high accuracy we have to take into account a Hilbert space of a dimension which is approximately 10 times the noise number $N_0$. It is therefore numerically challenging to implement the iterative procedure in cases where the noise number is large. If this is the case one should preferably work with the moments based maximum likelihood method presented in the previous section.

\section{Two channel detection and the positive $P$ distribution}
\begin{figure}[b]
\centering
\includegraphics[scale=.08]{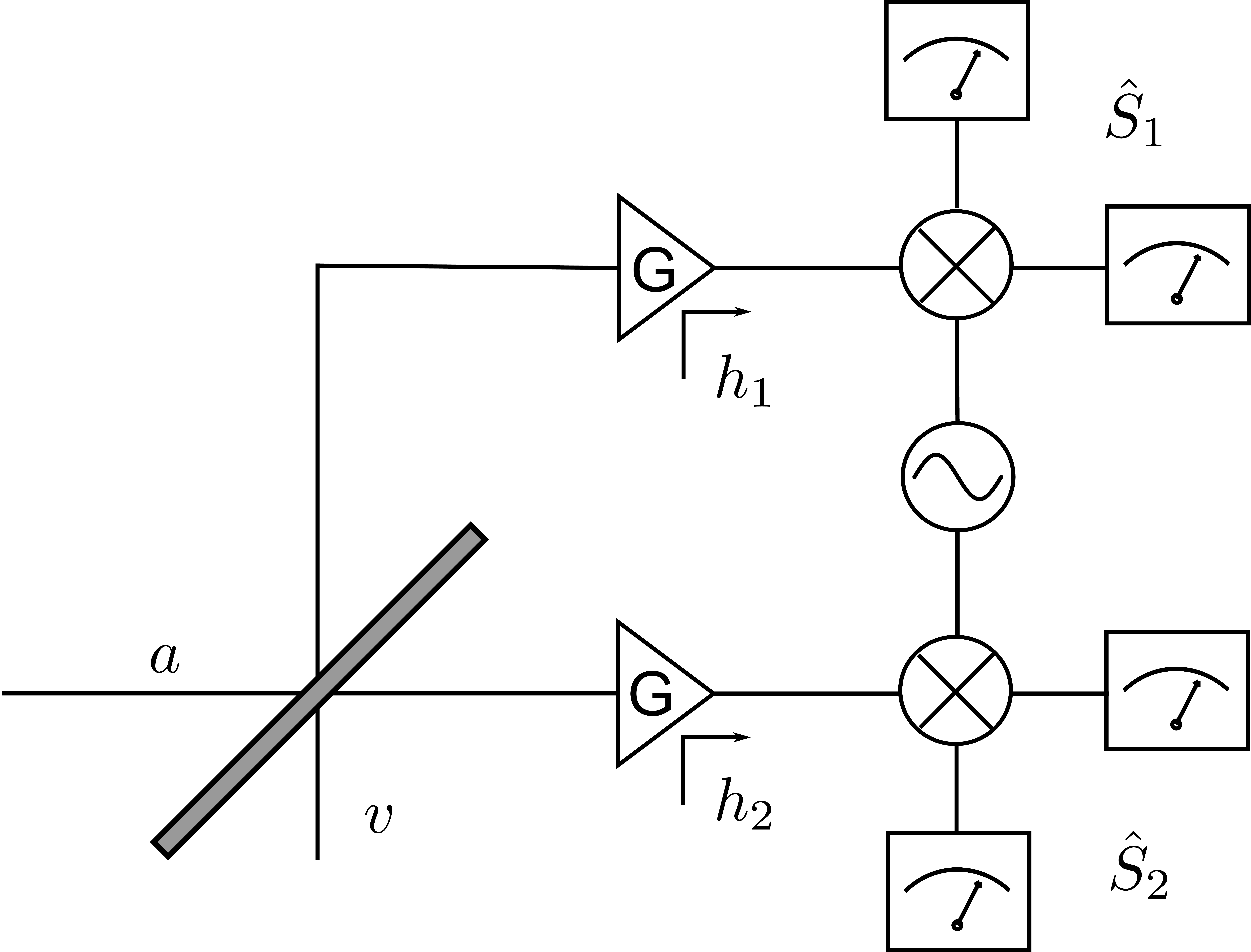}
\caption{Two channel detector with radiation incident from input mode $a$. Each of the beamsplitter outputs has an individual amplification stage adding noise in modes $h_1$ and $h_2$. In both channels the complex amplitude is measured \cite{daSilva2010,Bozyigit2011}.
}
\label{fig:twochanneldetection}
\end{figure}

We have already pointed out that field quadrature measurement is the most commonly used detection method for microwave frequency fields. Due to its well-established implementation it has also been possible to experimentally realize Hanbury Brown Twiss-type setups (see Fig.~\ref{fig:twochanneldetection}) where two instead of one complex amplitudes are measured \cite{Gabelli2004,Menzel2010,Bozyigit2011,Lang2011,Mariantoni2010}. The advantage of such a detection scheme is that ideally the system noise in the two detection arms is uncorrelated and only the signal mode $a$ contributes to the cross-correlations between the two output arms. In this section we provide a quantum optics description of a generic two-channel microwave detection chain \cite{Menzel2010,Mariantoni2010} as shown in \figref{fig:twochanneldetection}. We formulate the main advantages of such a measurement setup and show that under reasonable assumptions a direct measurement of a positive $P$ distribution  \cite{Drummond1980b} is realized. This relation between the positive $P$ distribution and the two-channel detection scheme gives important insight into the general statistical properties of the obtained measurement results.
\subsection{Two-channel detection}
The main difference between the one- and the two-channel setup depicted in \figref{fig:twochanneldetection} is the additional beamsplitter which splits the signal mode $a$ equally into two parts while introducing an additional mode $v$. As a result, the input modes at the two amplifiers are given by  $(a\pm v)/\sqrt{2}$ and the total measured complex amplitudes $\hat{S}_1$ and $\hat{S}_2$ can be expressed as
\begin{eqnarray}\label{eq:twoChannelSop}
\hat{S}_1  &=&  a + v + \sqrt{2} h_1^\dag,
\nonumber
\\
\hat{S}_2 &=& a - v + \sqrt{2} h_2^\dag,
\end{eqnarray}
where $h_1$ and $h_2$ are the modes accounting for the system  noise of the two detection chains. Under the assumptions justified below that (i) the mode $v$ is in the vacuum state, that (ii) all other modes are uncorrelated $\rho=\rho_{a}\otimes\rho_{h_1}\otimes\rho_{h_2}$, and that (iii) the noise has no phase-coherence $\langle h_2^m \rangle = 0 = \langle h_1^m \rangle  , \, \forall  m>0$ we find the following cross-correlations \cite{daSilva2010}
\begin{align}\label{eq:twoChannelMoments}
\langle (\hat{S}_1^\dag)^m \hat{S}^n_2 \rangle =
\langle (a^\dag)^m a^n \rangle.
\end{align}
The above assumptions require that (i)  the open beamsplitter port is connected to a bath of zero temperature, that (ii) the signal is not correlated with the two completely independent amplifier chains, and that (iii) the noise does not depend on the phase defined by the reference local oscillator, all of which can in good approximation be experimentally realized \cite{Bozyigit2011,Lang2011}

This result is remarkable since under realistic conditions the above cross-correlations completely describe the state of mode $a$ without any influence of the detector noise modes. This means that the scheme is largely independent of the choice of amplifiers and noise sources, even if the noise constitutes the majority of the power in the signals $S_1, S_2$.  As we show in the following these properties can be understood in terms of the positive $P$ function representation \cite{Drummond1980b}.
\subsection{The positive $P$ function}
The positive $P$ function was introduced by Drummond and Gardiner \cite{Drummond1980b} as a  theoretical concept for the solution of Fokker-Planck equations. In contrast to the Wigner function and Glauber-Sudarshan $P$ function it is completely positive and has all properties of a genuine probability distribution. The positive $P$ function  $P(\alpha, \beta^*)$  is defined as a non-diagonal expansion of the density matrix in the coherent state basis

\begin{align}\label{eq:DensMatposP}
\rho_a = \int_{\alpha,\beta}  P(\alpha, \beta) \frac{\ket{\alpha}\bra{\beta}}{\braket{\beta|\alpha}}.
\end{align}
Like the $P$ function it generates the normally ordered moments of the field operator
\begin{align}\label{eq:pospmoments}
\langle (a^\dag)^m a^n \rangle = \int_{\alpha,\beta}  P(\alpha, \beta) \alpha^n (\beta^*)^m.
\end{align}
Furthermore this four-dimensional probability distribution $P(\alpha, \beta^*)$ can be shown to be positive, not unique and to exist for any quantum state \cite{Drummond1980b}. To resolve the problem of uniqueness one can resort to the \emph{canonical choice} \cite{Braunstein1991}  of the positive $P$ function which is given by
\begin{align}
P_{\rm{can}}(\alpha,\beta) = \frac{1}{4\pi} \exp\left(-\frac{|\alpha-\beta|^2}{4} \right)  Q_a\left(\frac{\alpha + \beta}{2}\right).
\end{align}
While the positive $P$ function is often considered artificial and only of theoretical relevance, Braunstein~\textsl{et~al.} have shown that it can be interpreted as the probability distribution for the simultaneous measurement of four quadrature variables  \cite{Braunstein1991}. A scheme for an optical experiment was proposed by Agarwal \cite{Agarwal1994} based on fourfold balanced homodyne detection. To our knowledge this scheme has so far not been implemented, probably due to the significant experimental effort necessary.

\subsection{Two channel detection as a measurement of the positive $P$ function}
The scheme by Agarwal and the two channel microwave detection scheme in \figref{fig:twochanneldetection} are equivalent up to the presence of the amplifier noise. Furthermore under the assumptions made above about the noise,  the observables $\hat{S}_1, \hat{S}_2$ generate the normally ordered moments in the same way as the positive $P$ function. It is thus natural to assume that the probability distribution of the measurement data $P(S_1,S_2)$ \emph{is} a positive P-representation of the input mode $a$. In appendix \ref{app:psesf} we calculate this distribution and find
\begin{align}\label{eq:psesfresult}
P(S_1,S_2) = \frac{1}{4}\int_\beta P_a(\beta) Q_1\left(\frac{S^*_1-\beta^*}{\sqrt{2}}\right) Q_2\left(\frac{S^*_2 - \beta^*}{\sqrt{2}}\right) .
\end{align}
where $Q_{1,2}(\alpha)$ are the $Q$ functions of the system noise modes $h_1, h_2$. When the noise added to both channels is in a thermal state with mean photon number $N_0$ Eq.~(\ref{eq:psesfresult}) simplifies to
\begin{align}\label{eq:posPwithNoise}
P(S_1,S_2) &= \frac{\exp\left(-\frac{|S_1-S_2|^2}{4(N_0+1)}\right) }{4 \pi(N_0+1)} \, W_a\left(\frac{S_1+S_2}{2}, -1 - 2 N_0\right)
\end{align}
and for quantum limited detection, i.e. $N_0 = 0$, to
\begin{align}
P(S_1,S_2) &= \frac{1}{4\pi}\exp\left(-\frac{|S_1-S_2|^2}{4} \right)  Q_a\left(\frac{S_1+S_2}{2}\right)\\
&= P_{\rm{can}} \left( S_1, S_2\right).
\end{align}
The compelling result is that for quantum limited detection the measurement data distribution corresponds to the canonical choice of the positive P-representation $P_{\rm{can}} (\alpha,\beta)$. Moreover, we show in Appendix \ref{app:proofPosP} that the measured distribution is always a positive $P$ distribution
\begin{equation}
\frac{1}{\pi}\int_{S_1,S_2} \,  P(S_1,S_2) \frac{\langle \alpha |S_1 \rangle\langle S_2|\alpha \rangle}{\langle S_2|S_1 \rangle} =  Q_a(\alpha)
\label{eq:toShow}
\end{equation}
for any thermal populations $N_1$, $N_2$ in the detector noise modes. As a consequence, the density matrix can be directly evaluated from the measured $P(S_1,S_2)$ using Eq.~(\ref{eq:DensMatposP}) even in the presence of significant thermal noise of unequal powers in the detection chains. These results suggest that the measurement of a positive $P$ distribution is possible with current microwave frequency quadrature detection setups such as the one described in Ref.~\cite{Bozyigit2011}.
\section{Joint Tomography Scheme for a Qubit - Photon Field System}
In the previous sections we have described experimental schemes for characterizing microwave radiation on a quantum level. In many experiments where such radiation fields are relevant they interact with stationary quantum objects such as superconducting qubits. Similar to optical photons in atomic systems \cite{Blinov2004,Volz2006,Moehring2007,Togan2010,Kimble2008,Ritter2012,Stute2012}, itinerant quantum microwave fields have the potential to act as a quantum channel to connect spatially separated superconducting circuits with each other.
In this section we discuss a state tomography scheme to reconstruct the joint density matrix of the qubit-photon field system \cite{Wallentowitz1997} using linear detection. The scheme is an extension of the photon field tomography methods described in the previous sections and is applicable to any system in which the complex field amplitude $\hat{S}$ and the qubit state population along an arbitrary axis can both be measured in each trial of an experimental run. The presented method allows to include finite detection efficiencies for both the qubit and the photon field detection.
\subsection{Qubit state tomography}
In order to describe the joint tomography scheme we first discuss the concepts of qubit tomography. For the reconstruction of the qubit density matrix $\rho_{\sigma}$ one measures the Pauli expectation values $\langle \sigma_i \rangle$  along the different spin axes $\sigma_i \in \{\sigma_x,\sigma_y,\sigma_z\}$ which in the measurement basis $\{|0_z\rangle,|1_z\rangle\}$ are represented by the corresponding Pauli matrices.
After state preparation the qubit is rotated such that the desired spin component $\sigma_i$ points along the measurement axis. This rotation is followed by a read-out procedure during which the measurement result is encoded in a classical quantity $q$ \cite{Wallraff2005}. In the context of circuit QED  single shot read-out \cite{Mallet2009,Vijay2011} is not always available and $q$ can take a continuous spectrum of values where depending on the qubit state each value has a probability $D_{{i}}(q)$ to occur. Here the index $i \in \{x,y,z\}$ specifies the measurement basis.

The distribution $D_{{i}}(q)$ obtained after repeating the measurement many times can be fitted to the weighted sum of 2 reference distributions $p_0(q)$ for the ground and $p_1(q)$ for the excited state
\begin{equation}
D_i(q) \doteq \frac{1- \langle\sigma_i\rangle}{2} p_1(q) + \frac{1+\langle\sigma_i\rangle}{2} p_0(q)
\label{eq:extractQubitPop}
\end{equation}
to extract the Pauli expectation values $\langle\sigma_i\rangle$. Based on these values the density matrix can then be determined as $\rho_\sigma = \frac{1}{2}(\mathds{1} + \sum_{i } \langle\sigma_i\rangle \sigma_i )$. Note that instead of using Eq.~(\ref{eq:extractQubitPop}) the qubit population can also be extracted from the mean values of $q$ \cite{Bianchetti2009} as done in many experiments .
\subsection{Joint tomography}
A joint tomography scheme is expected to allow for a full characterization of the system also when photon field and qubit are correlated with each other. The goal is to determine all matrix elements $\langle s,n|\rho_{\sigma,a}|s^\prime,m \rangle$ of the joint density matrix $\rho_{\sigma,a}$ where $s,s^\prime \in \{0_z,1_z\}$ label the qubit basis states and $n,m$ the photon field number states. It can be shown that these matrix elements are uniquely determined by the set of moments $\langle(a^\dagger)^n a^m {\sigma}_i\rangle $ in the following way
\begin{eqnarray}
\langle 0_z,m|\rho_{\sigma,a}|0_z,n \rangle &=& \frac{1}{2}\mathcal{M}\left(\langle(a^\dagger)^n a^m\rangle + \langle(a^\dagger)^n a^m \sigma_z\rangle \right)
\nonumber
\\
\langle 1_z,m|\rho_{\sigma,a}|1_z,n \rangle &=& \frac{1}{2}\mathcal{M}\left(\langle(a^\dagger)^n a^m\rangle - \langle(a^\dagger)^n a^m \sigma_z\rangle \right)
\nonumber
\\
\langle 1_z,m|\rho_{\sigma,a}|0_z,n \rangle&=& \frac{1}{2}\mathcal{M}\left(\langle(a^\dagger)^n a^m(\sigma_x + i \sigma_y)\rangle \right)
\label{eq:densityMatrixRec}
\end{eqnarray}
Here, $\mathcal{M}$ is the linear map from the moments to the density matrix in the number state basis as defined in Eq.~(\ref{eq:momsToDensMat}). The scheme described in the following allows to measure all the necessary moments in Eq.~(\ref{eq:densityMatrixRec}).

We consider the case that in each trial of an experimental run both $q$, characterizing the qubit state, and the complex amplitude $\hat{S}$, characterizing the photon field, are measured. For each state preparation both numbers are stored in a 3D histogram $D_i(S,q)$ where the index $i$ labels the chosen qubit rotation before measurement. To evaluate the desired expectation values we first determine the Pauli expectation values $\langle\sigma_i \rangle_S$ conditioned on the complex amplitude result $S$. This is done by fitting each trace of  $D_i(S,q)$ along the $q$ axes to the calibration histograms $p_0(q)$ and $p_1(q)$. Based on the knowledge of $\langle\sigma_i \rangle_S$ we determine the photon field distributions conditioned on a specific qubit measurement result as
\begin{eqnarray}
D_{0_i}^{}(S) &=&  \mathcal{N}_0\frac{1+\langle\sigma_i\rangle_{S}}{2} \sum_{q}{}D_i^{}(S,q)
\nonumber
\\
D_{1_i}^{}(S) &=& \mathcal{N}_1\frac{1-\langle\sigma_i\rangle_{S}}{2} \sum_{q}{}D_i^{}(S,q)
\label{eq:conditionedQPD}
\end{eqnarray}
where $\mathcal{N}_0$ and $\mathcal{N}_1$ are appropriate normalization constants which guarantee that $\int_S D_{0_i}^{}(S) = \int_S D_{1_i}^{}(S)=1$. For example, $D_{0_x}(S)$ is the photon field distribution under the condition that the qubit is measured with result 0 in the $x$-basis.

Given these histograms one can evaluate the conditioned moments $\langle (\hat{S}^\dagger)^n \hat{S}^m \rangle|0_i$ and $\langle (\hat{S}^\dagger)^n \hat{S}^m \rangle|1_i$ using Eq.~(\ref{eq:moments}).
If signal field and qubit are not correlated with the noise $\rho_{\sigma,a} \otimes \rho_h$ we can use the techniques described in section \ref{sec:photonTomo} to extract the desired quantities $\langle(a^\dagger)^n a^m \sigma_i\rangle$. In recent experiments we have used the presented method to analyze  correlations between single itinerant microwave photons entangled with a superconducting qubit \cite{Eichler2012a}.
\subsection{Maximum-likelihood state estimation for a joint system}
As for photon state tomography, it is desirable to find a set of POVM operators for the described measurement scheme. This allows to construct an iterative maximum-likelihood state estimation procedure and furthermore provides insight into the conditioned histograms introduced in Eq.~(\ref{eq:conditionedQPD}).

For a perfect single shot qubit readout with a binary measurement result $0$ or $1$, the POVMs are given by the projectors onto eigenstates of the Pauli operators, $\hat{\Pi}_{0_i} = |{0_i}\rangle\langle{0_i}|$ and $\hat{\Pi}_{1_i}=|{1_i}\rangle\langle{1_i}|$. They are complete in the sense that an arbitrary qubit density matrix can be explicitly written as a linear combination of  projectors $\rho_\sigma= \frac{1}{2}(\mathds{1} + \sum_{i } \langle\sigma_i\rangle (\hat{\Pi}_{0_i} - \hat{\Pi}_{1_i}))$.

Including the photon field measurement the total set of POVMs is
\begin{equation}
\hat{\Pi}_{\alpha,s_i}^{} =
\hat{\Pi}_{\alpha}^{[\rho_h]} \otimes \hat{\Pi}_{s_i}
\label{eq:POVMQubitPhoton}
\end{equation}
with $s \in \{0,1\}$ and $i \in \{x,y,z\}$ for which the expectation values with respect to the total density matrix are related to the measured histograms by
\begin{eqnarray}
{\rm Tr}[\rho_{\sigma,a} \hat{\Pi}_{\alpha,0_i}^{}] &\doteq& \frac{1+\langle \sigma_i \rangle}{2} D_{0_i}^{}(\alpha)
\nonumber
\\
{\rm Tr}[\rho_{\sigma,a}\hat{\Pi}_{\alpha,1_i}^{}] &\doteq&  \frac{1-\langle \sigma_i \rangle}{2} D_{1_i}^{}(\alpha)
\label{eq:POVMMeasurementRelation}
\end{eqnarray}
with $\langle \sigma_i \rangle$ being the unconditioned Pauli expectation values. Note that this remains valid for qubit readout with limited single shot fidelity since the storage of the data in 3-dimensional histograms allows for capturing all the necessary qubit-photon correlations. By fitting the 3D histogram data along the $q$-axes to the expected ground and excited state distributions (see Eq.~(\ref{eq:extractQubitPop})) we account for the finite read-out efficiency. Using the set of POVMs given in Eq.~(\ref{eq:POVMQubitPhoton}) we are able to use the iterative maximum likelihood procedure described above to estimate the most likely density matrix for the combined system.
\section{Conclusions}
In summary, we have presented schemes for the characterization of a single radiation field mode and its entanglement with a two-level system based on linear amplification and quadrature detection. We have discussed single channel and two-channel detection schemes and showed that the latter enables a direct measurement of a positive $P$ function even in the presence of significant added amplifier noise. For both the photon field and the joint tomography scheme we have discussed maximum-likelihood procedures which take into account the full measured quasi-probability distributions.

Due to the recent progress in the development of quantum-limited amplifiers and microwave photon counters we believe that  the investigation of  itinerant microwave photons has a great potential in quantum science. The state tomography methods described in this paper enable the experimental characterization of such microwave radiation fields and of optical fields detected with finite efficiency.
\begin{acknowledgments}
The authors would like to acknowledge fruitful discussions with Gerard Milburn, Marcus~P. da Silva, Barry Sanders, and Alexandre Blais. This work was supported by the European Research Council (ERC) through a Starting Grant and by ETHZ.
\end{acknowledgments}
\appendix
\section{Temporal mode matching}\label{app:temporalmodematching}
Here we discuss the relation between the single \emph{time-independent} mode $a$ describing the photon pulse which is to be characterized by state tomography and the \emph{time-dependent}  field $a_{\rm out}(t)$  which is continuously sampled in an experiment.  The link between the two is given by a mode-matching relation
\begin{equation}a = \int \text{d}t f(t) a_{\rm out}(t)
\end{equation}
where the normalization condition $\int \text{d}t |f(t)|^2 =1$  of the temporal profile function $f(t)$ guarantees that $[a,a^\dag] =1$ is satisfied. The best choice of $f(t)$ depends on the temporal shape of the field, i.e. the properties of the  coupling between radiation source and the bath under observation. In the following we discuss optimal temporal mode matching for a single-sided cavity acting as the radiation source.

We represent the cavity mode by the creation operator $A(t)$, which in a Heisenberg picture is time-dependent. We assume that at time $t=0$ the cavity is prepared in a specific state described by the statistics of $A(0)$ and then left under free evolution \cite{Walls1994}
\begin{equation}
A(t) = e^{-\frac{\kappa t}{2}} A(0) + \sqrt{\kappa} e^{- \frac{\kappa t}{2}}\int_0^{t} \text{d}\tau e^{\frac{\kappa \tau}{2}} a_{\text{in}}(\tau).
\label{eq:freeEvolution}
\end{equation}
From input-output theory \cite{Gardiner1985} we know that the cavity field decays with rate $\kappa$ into the output modes according to
\begin{align}
\begin{split}
a_{\rm out}(t) &= \sqrt{\kappa} A(t) - a_{\rm{in}}(t).
\end{split}
\end{align}
The input modes $a_{\rm in}(t)$  can be understood as a continuous stream of independent modes each reaching the resonator at time $t$ and ideally carrying only the vacuum noise.

By inserting the above expressions into the definition of $a$ we get only one term $A(0)\,\sqrt{\kappa} \int_{0}^\infty \text{d}t e^{-\kappa t}$ depending on the cavity field. In order to maximize the efficiency in detecting the state prepared at time $t=0$ we have to find $f(t)$ which maximizes this term. The choice $f(t) = \sqrt{\kappa} e^{-\frac{\kappa t}{2}} \Theta(t)$ does so, where $\Theta(t)$ is the Heaviside step function.
The total expression then reduces to
\begin{align}
\begin{split}\label{eq:AoneSideCavity}
a &= A(0)  \kappa\int_{0}^{\infty}\text{d}t  e^{-\kappa t} - \kappa^{1/2}  \int_0^{\infty} e^{-\frac{\kappa t}{2}} a_{\rm{in}}(t) {\rm d}t\\
  &+ \kappa^{3/2}  \int_0^{\infty} e^{-\kappa t} \int_0^t e^{\frac{\kappa \tau}{2}} a_{\rm{in}}(\tau) {\rm d}\tau {\rm d}t.
\end{split}
\end{align}
which due to the identity
\begin{align}
\begin{split}
 & \kappa^{3/2}  \int_0^{\infty} e^{-\kappa t} \int_0^t e^{\frac{\kappa \tau}{2}} a_{\rm{in}}(\tau) {\rm d}\tau {\rm d}t \\
 = & \kappa^{3/2}  \int_0^{\infty} \left( \int_0^{\infty} \Theta(t-\tau) e^{-\kappa t} {\rm d}t\right) e^{\frac{\kappa \tau}{2}} a_{\rm{in}}(\tau) {\rm d}\tau \\
 = & \kappa^{1/2}  \int_0^{\infty} e^{-\frac{\kappa \tau}{2}} a_{\rm{in}}(\tau) {\rm d}\tau.
\end{split}
\end{align}
simplifies to
\begin{align}
a = A(0).
\end{align}
By proper choice of $f(t)$ we can thus recover the state of the source field $A(0)$ with unit efficiency in the transmission line. Note that a finite mode matching efficiency only reduces the total detection efficiency but does not affect the statistical properties of $a$.
\section{Probability distribution for two channel complex envelopes}\label{app:psesf}
Here we calculate the joint probability distribution of the complex envelopes in a two channel detection scheme along the lines of Ref.~\cite{Agarwal1994}. By definition the probability distribution of the measurement data $S_1,S_2$ is given by the Fourier transform of the characteristic function
\begin{align}\label{eq:sefProbDistDef}
P(S_1,S_2) = \frac{1}{\pi^4} \int_{z_1,z_2} e^{z^*_1 S_1 + z^*_2 S_2 - z_1 S^*_1 - z_2 S^*_2} \chi_{SS} (z_1,z_2).
\end{align}
where
\begin{align}
\chi_{SS} (z_1,z_2) = \left\langle  e^{z_1\hat{S}^\dag_1 + z_2\hat{S}^\dag_2 } e^{-z^*_1\hat{S}_1 -z^*_2\hat{S}_2 } \right\rangle.
\end{align}
By substituting the operator Eqs.~(\ref{eq:twoChannelSop}) for $\hat{S}_1$ and $\hat{S}_2$ we find
\begin{eqnarray}
\chi_{SS} (z_1,z_2) &=&
\\
& &\hspace{-20mm} \chi_a({z_1+z_2}) \chi_v({z_1-z_2}) \chi_{h_1}(-\sqrt{2}z_1^*) \chi_{h_2}(-\sqrt{2}z_2^*)
\nonumber
\end{eqnarray}
where we introduced the characteristic functions for the four different modes as
\begin{align}
\chi_a(z) &= \langle e^{z a^\dagger} e^{-z^* a} \rangle &= \int_\beta P_a(\beta) e^{z \beta^* - z^* \beta} \\
\chi_v(z) &= \langle e^{z v^\dagger} e^{-z^* v} \rangle\\
\chi_{h_i}(z) &= \langle  e^{-z^* h_i} e^{z h_i^\dagger} \rangle &= \int_\beta Q_i(\beta) e^{z \beta^* - z^* \beta}
\end{align}

We can simplify these expressions by introducing the following physical assumptions. First, mode $v$ is assumed to be in the vacuum state. In this case its characteristic function is the identity and we have
\begin{align}
\chi_{SS} (z_1,z_2) = \chi_a({z_1+z_2}) \chi_{h_1}(- \sqrt{2} z_1^*) \chi_{h_2}(- \sqrt{2} z_2^*) .
\end{align}
Substituting this equation and the integral forms of the characteristic functions in the definition of $P(S_1,S_2)$ we get
\begin{align}
P(S_1,S_2) = \int_{\beta,\eta,\gamma} P_a(\beta) Q_1(\eta) Q_2(\gamma)\,D,
\label{eq:positvePThermal}
\end{align}
where
\begin{eqnarray}
D &=& \pi^{-4} \int_{z_1} \exp({z_1 (\sqrt{2} \eta + {\beta^*} - S^*_1) - \text{c.c.} )})
\nonumber
\\
&&\hspace{10mm}\int_{z_2}  \exp({z_2 (\sqrt{2} \gamma + {\beta^*} - S^*_2) - \text{c.c.})}).
\nonumber
\\
&=& \frac{1}{4}\delta(\eta^* + \frac{\beta - S_1}{\sqrt{2}}) \delta(\gamma^* + \frac{\beta - S_2}{\sqrt{2}}),
\end{eqnarray}
which reduces Eq.~(\ref{eq:positvePThermal}) to
\begin{equation}
P(S_1,S_2) = \frac{1}{4} \int_\beta P_a(\beta) Q_1\left(\frac{S^*_1 -  \beta^*}{\sqrt{2}}\right) Q_2\left( \frac{S^*_2 - \beta^*}{\sqrt{2}}\right).
\end{equation}
\subsection{Thermal noise}
In the next step we assume that the noise modes $h_1, h_2$ are in thermal states $e^{-|\alpha|^2/N_i+1}/\pi (N_i + 1)$ \cite{Cahill1969} with mean photon numbers $N_1, N_2$. The probability distribution of the measurement data is then
\begin{align}
P(S_1,S_2) =\int_\beta P_a(\beta)\,\frac{\exp\left({ -\frac{|S_1 - {\beta}|^2}{2(N_1+1)}  -\frac{|S_2 - \beta|^2}{2(N_2+1)}}\right)}{4 \pi^2 (N_1+1)(N_2+1)}  .
\label{eq:PpositiveOrig}
\end{align}
Comparing this with the formula for the s-parametrized QPD
\begin{align}
W_a(\bar{S},s)& = \frac{2 \pi^{-1}}{1-s} \int_\beta P_a(\beta) \exp\left(-\frac{2 |\bar{S} - \beta|^2}{1-s}\right)
\end{align}
we can identify the relation
\begin{align}
P(S_1,S_2) = \frac{1}{2 \pi N_{\rm{tot}}} e^{-\frac{|S_1 - S_2|^2}{2 N_{\rm{tot}}}} \, W_a(\bar{S},s)
\label{eq:Ppositive}
\end{align}
by defining
\begin{align}
\bar{S} &= \frac{N_1+1}{N_{\rm{tot}}} S_1 + \frac{N_2+1}{N_{\rm{tot}}} S_2, \\
s &= -1 - \frac{4 \, N_{1} N_{2} +2\, N_{1} + 2\,N_{2} }{N_{\rm{tot}}},\\
N_{\rm{tot}} &= N_1 + N_2 + 2\,.
\end{align}
If we have the same noise level on both channels $N_1 = N_2 = N_0$, we find $s = -1 - 2 N_0$ and thus
\begin{align}
P(S_1,S_2) &= \frac{e^{-\frac{|S_1-S_2|^2}{4(N_0+1)}} }{4 \pi({N_0+1})} \, W_a\left(\frac{S_1+S_2}{2}, -1 - 2 N_0\right)
\end{align}
In the case of quantum limited detection $N_1 = N_2 = 0$ we have $s = -1$ and the distribution of our measurement data corresponds to the canonical positive P-representation of mode $a$
\begin{align}
P(S_1,S_2) &= \frac{1}{4 \pi} e^{-\frac{|S_1-S_2|^2}{4}}  Q_a\left(\frac{S_1+S_2}{2}\right).
\end{align}
\subsection{Equivalence to canonical positive $P$ function}\label{app:proofPosP}
To proof that Eq.~(\ref{eq:Ppositive}) is also a positive $P$ function when the thermal noise levels are unequal $N_1 \neq N_2$ we show that \cite{Braunstein1991}
\begin{equation}
\frac{1}{\pi}\int_{S_1,S_2} \,  P(S_1,S_2) \frac{\langle \alpha |S_1 \rangle\langle S_2|\alpha \rangle}{\langle S_2|S_1 \rangle}\doteq Q(\alpha).
\label{eq:toShow}
\end{equation}
Using Eq.~(\ref{eq:PpositiveOrig}), the lhs of Eq.~({\ref{eq:toShow}}) is
\begin{eqnarray}
\int_{\beta,S_1,S_2} P_a(\beta)\,\frac{\exp\left({ -\frac{|S_1 - {\beta}|^2}{2(N_1+1)}  -\frac{|S_2 - \beta|^2}{2(N_2+1)}}\right)}{4 \pi^3 (N_1+1)(N_2+1)}
\frac{\langle \alpha |S_1 \rangle\langle S_2|\alpha \rangle}{\langle S_2|S_1 \rangle}
\nonumber
\\
\end{eqnarray}
This is a multi-dimensional Gaussian integral in the variables $S_1, S_2$ and can be solved to give
\begin{equation}
\frac{1}{\pi}\int_{\beta} P_a(\beta) e^{-|\beta - \alpha|^2},
\end{equation}
which is exactly the $Q$ function.
\bibliographystyle{apsrevshort}


\begin{thebibliography}{10}

\bibitem{Deleglise2008}
S.~Deleglise \emph{et~al.}, Nature, \textbf{455}, 510 (2008).

\bibitem{Hofheinz2009}
M.~Hofheinz \emph{et~al.}, Nature, \textbf{459}, 546 (2009).

\bibitem{Brune1996}
M.~Brune \emph{et~al.}, Phys. Rev. Lett., \textbf{76}, 1800 (1996).

\bibitem{Wang2009a}
H.~Wang \emph{et~al.}, Phys. Rev. Lett., \textbf{103}, 200404 (2009).

\bibitem{Wang2011b}
H.~Wang \emph{et~al.}, Phys. Rev. Lett., \textbf{106}, 060401 (2011).

\bibitem{Sayrin2011}
C.~Sayrin \emph{et~al.}, Nature, \textbf{477}, 73 (2011).

\bibitem{Castellanos2008}
M.~A. Castellanos-Beltran \emph{et~al.}, Nat. Phys., \textbf{4}, 929 (2008).

\bibitem{Bergeal2012}
N.~Bergeal, F.~Schackert, L.~Frunzio, and M.~H. Devoret, Phys. Rev. Lett.,
  \textbf{108}, 123902 (2012).

\bibitem{Houck2007}
A.~Houck \emph{et~al.}, Nature, \textbf{449}, 328 (2007).

\bibitem{Astafiev2010}
O.~Astafiev \emph{et~al.}, Science, \textbf{327}, 840 (2010).

\bibitem{Bozyigit2011}
D.~Bozyigit \emph{et~al.}, Nature Physics, \textbf{7}, 154 (2011).

\bibitem{Lang2011}
C.~Lang \emph{et~al.}, Physical Review Letters, \textbf{106}, 243601 (2011).

\bibitem{Mallet2011}
F.~Mallet \emph{et~al.}, Phys. Rev. Lett., \textbf{106}, 220502 (2011).

\bibitem{Eichler2011}
C.~Eichler \emph{et~al.}, Physical Review Letters, \textbf{106}, 220503 (2011).

\bibitem{Menzel2010}
E.~P. Menzel \emph{et~al.}, Phys. Rev. Lett., \textbf{105}, 100401 (2010).

\bibitem{Eichler2011a}
C.~Eichler \emph{et~al.}, Physical Review Letters, \textbf{107}, 113601 (2011).

\bibitem{Flurin2012}
E.~Flurin \emph{et~al.}, arXiv:1204.0732v1 (2012).

\bibitem{Yurke2006}
B.~Yurke and E.~Buks, J. Lightwave Technol., \textbf{24}, 5054 (2006).

\bibitem{Yurke1987}
B.~Yurke, J. Opt. Soc. Am. B, \textbf{4}, 1551 (1987).

\bibitem{Bergeal2010}
N.~Bergeal \emph{et~al.}, Nature, \textbf{465}, 64 (2010).

\bibitem{Kinion2008}
D.~Kinion and J.~Clarke, Applied Physics Letters, \textbf{92}, 172503 (2008).

\bibitem{Hatridge2011}
M.~Hatridge \emph{et~al.}, Phys. Rev. B, \textbf{83}, 134501 (2011).

\bibitem{Chen2011a}
Y.-F. Chen \emph{et~al.}, Phys. Rev. Lett., \textbf{107}, 217401 (2011).

\bibitem{Romero2009}
G.~Romero, J.~J. Garc\'\i{}a-Ripoll, and E.~Solano, Phys. Rev. Lett.,
  \textbf{102}, 173602 (2009).

\bibitem{Frey2012}
T.~Frey \emph{et~al.}, Physical Review Letters, \textbf{108}, 046807 (2012).

\bibitem{Delbecq2011}
M.~R. Delbecq \emph{et~al.}, Phys. Rev. Lett., \textbf{107}, 256804 (2011).

\bibitem{Schuster2010}
D.~I. Schuster \emph{et~al.}, Phys. Rev. Lett., \textbf{105}, 040503 (2010).

\bibitem{Kubo2010}
Y.~Kubo \emph{et~al.}, Phys. Rev. Lett., \textbf{105}, 140502 (2010).

\bibitem{Hogan2012}
S.~D. Hogan \emph{et~al.}, Physical Review Letters, \textbf{108}, 063004
  (2012).

\bibitem{Agarwal1994}
G.~S. Agarwal and S.~Chaturvedi, Phys. Rev. A, \textbf{49}, R665 (1994).

\bibitem{Carmichael2008Book}
H.~J. Carmichael, \emph{Statistical Methods in Quantum Optics 2: Non-Classical
  Fields} (Springer-Verlag, 2008).

\bibitem{Lvovsky2009}
A.~I. Lvovsky and M.~G. Raymer, Rev. Mod. Phys., \textbf{81}, 299 (2009).

\bibitem{Scully1997}
M.~O. Scully and M.~S. Zubairy, \emph{Quantum Optics} (Cambridge University
  Press, 1997).

\bibitem{Arthurs1965}
E.~Arthurs and J.~L. Kelly, Bell. Syst. Tech. J., \textbf{44}, 725 (1965).

\bibitem{Braunstein1991}
S.~L. Braunstein, C.~M. Caves, and G.~J. Milburn, Phys. Rev. A, \textbf{43},
  1153 (1991).

\bibitem{Caves1994}
C.~M. Caves and P.~D. Drummond, Rev. Mod. Phys., \textbf{66}, 481 (1994).

\bibitem{Welsch1999}
D.-G. Welsch, W.~Vogel, and T.~Opatrn? (Elsevier, 1999), vol.~39 of
  \emph{Progress in Optics}, 63 -- 211.

\bibitem{Yuen1980}
H.~Yuen and J.~Shapiro, Information Theory, IEEE Transactions on, \textbf{26},
  78  (1980).

\bibitem{Noh1991}
J.~W. Noh, A.~Foug\`eres, and L.~Mandel, Phys. Rev. Lett., \textbf{67}, 1426
  (1991).

\bibitem{Haus1962}
H.~A. Haus and J.~A. Mullen, Phys. Rev., \textbf{128}, 2407 (1962).

\bibitem{Caves1982}
C.~M. Caves, Phys. Rev. D, \textbf{26}, 1817 (1982).

\bibitem{Clerk2010}
A.~A. Clerk \emph{et~al.}, Rev. Mod. Phys., \textbf{82}, 1155 (2010).

\bibitem{daSilva2010}
M.~P. da~Silva, D.~Bozyigit, A.~Wallraff, and A.~Blais, Physical Review A,
  \textbf{82}, 043804 (2010).

\bibitem{Leonhardt1994}
U.~Leonhardt and H.~Paul, Phys. Rev. Lett., \textbf{72}, 4086 (1994).

\bibitem{Loudon2000}
R.~Loudon, \emph{The Quantum Theory of Light} (Oxford U, 2000).

\bibitem{Filippov2011}
S.~N. Filippov and V.~I. Man'ko, Phys. Rev. A, \textbf{84}, 033827 (2011).

\bibitem{Cahill1969a}
K.~E. Cahill and R.~J. Glauber, Phys. Rev., \textbf{177}, 1882 (1969).

\bibitem{Carmichael1999Book}
H.~J. Carmichael, \emph{Statistical Methods in Quantum Optics 1: Master
  Equations and Fokker-Planck Equations} (Springer-Verlag, 1999).

\bibitem{Haroche2006}
S.~Haroche and J.-M. Raimond, \emph{Exploring the Quantum: Atoms, Cavities, and
  Photons} (OUP Oxford, 2006).

\bibitem{Glauber1963c}
R.~J. Glauber, Phys. Rev., \textbf{131}, 2766 (1963).

\bibitem{Kim1997}
M.~S. Kim, Phys. Rev. A, \textbf{56}, 3175 (1997).

\bibitem{Vijay2011}
R.~Vijay, D.~H. Slichter, and I.~Siddiqi, Phys. Rev. Lett., \textbf{106},
  110502 (2011).

\bibitem{Leonhardt1993}
U.~Leonhardt and H.~Paul, Phys. Rev. A, \textbf{48}, 4598 (1993).

\bibitem{Vogel2000}
W.~Vogel, Phys. Rev. Lett., \textbf{84}, 1849 (2000).

\bibitem{Kiesel2011}
T.~Kiesel, W.~Vogel, B.~Hage, and R.~Schnabel, Phys. Rev. Lett., \textbf{107},
  113604 (2011).

\bibitem{Buzek1996}
V.~Bu\v{z}ek, G.~Adam, and G.~Drobn\'y, Phys. Rev. A, \textbf{54}, 804 (1996).

\bibitem{Fink2010}
J.~M. Fink \emph{et~al.}, Physical Review Letters, \textbf{105}, 163601 (2010).

\bibitem{Herzog1996}
U.~Herzog, Phys. Rev. A, \textbf{53}, 2889 (1996).

\bibitem{Chow2012}
J.~M. Chow \emph{et~al.}, arXiv:1202.5344 (2012).

\bibitem{Vandenberghe1996}
L.~Vandenberghe and S.~Boyd, SIAM Review, \textbf{38}, 49 (1996).

\bibitem{Eichler2012a}
C.~Eichler \emph{et~al.}, in preparation (2012).

\bibitem{Guta2012}
M.~Guta, T.~Kypraios, and I.~Dryden, arXiv:1206.4032v1 (2012).

\bibitem{Rehacek2008}
J.~Rehácek, D.~Mogilevtsev, and Z.~Hradil, New Journal of Physics,
  \textbf{10}, 043022 (2008).

\bibitem{Audenaert2009}
K.~M.~R. Audenaert and S.~Scheel, New Journal of Physics, \textbf{11}, 023028
  (2009).

\bibitem{Dobek2011}
K.~Dobek \emph{et~al.}, Phys. Rev. Lett., \textbf{106}, 030501 (2011).

\bibitem{Nielsen2000}
M.~A. Nielsen and I.~L. Chuang, \emph{Quantum Computation and Quantum
  Information} (Cambridge University Press, 2000).

\bibitem{Hradil2004}
Z.~Hradil, J.~Rehacek, J.~Fiurasek, and M.~Jezek, in \emph{Quantum State
  Estimation}, edited by M.~Paris and J.~Rehacek (Springer, 2004), Lecture
  Notes in Physics.

\bibitem{Lvovsky2004}
A.~I. Lvovsky, Journal of Optics B: Quantum and Semiclassical Optics,
  \textbf{6}, S556 (2004).

\bibitem{Rehacek2007}
J.~\ifmmode \check{R}\else \v{R}\fi{}eh\'a\ifmmode~\check{c}\else \v{c}\fi{}ek,
  Z.~c.~v. Hradil, E.~Knill, and A.~I. Lvovsky, Phys. Rev. A, \textbf{75},
  042108 (2007).

\bibitem{Hradil1997}
Z.~Hradil, Phys. Rev. A, \textbf{55}, R1561 (1997).

\bibitem{Hradil2006}
Z.~Hradil, D.~Mogilevtsev, and J.~\v{R}eh\'{a}\v{c}ek, Phys. Rev. Lett.,
  \textbf{96}, 230401 (2006).

\bibitem{Mogilevtsev2007}
D.~Mogilevtsev, J.~\ifmmode \check{R}\else
  \v{R}\fi{}eh\'a\ifmmode~\check{c}\else \v{c}\fi{}ek, and Z.~Hradil, Phys.
  Rev. A, \textbf{75}, 012112 (2007).

\bibitem{Babichev2004}
S.~A. Babichev, J.~Appel, and A.~I. Lvovsky, Phys. Rev. Lett., \textbf{92},
  193601 (2004).

\bibitem{Tipsmark2011}
A.~Tipsmark \emph{et~al.}, Phys. Rev. A, \textbf{84}, 050301 (2011).

\bibitem{Usuga2010}
M.~A. Usuga \emph{et~al.}, Nat Phys, \textbf{6}, 767 (2010).

\bibitem{Helstrom1976Book}
C.~W. Helstrom, \emph{Quantum Detection and Estimation Theory} (Academic Press,
  New York, 1976).

\bibitem{Wuensche1991}
A.~Wunsche, Quantum Optics: Journal of the European Optical Society Part B,
  \textbf{3}, 359 (1991).

\bibitem{Gabelli2004}
J.~Gabelli \emph{et~al.}, Phys. Rev. Lett., \textbf{93}, 056801 (2004).

\bibitem{Mariantoni2010}
M.~Mariantoni \emph{et~al.}, Phys. Rev. Lett., \textbf{105}, 133601 (2010).

\bibitem{Drummond1980b}
P.~D. {Drummond} and C.~W. {Gardiner}, Journal of Physics A Mathematical
  General, \textbf{13}, 2353 (1980).

\bibitem{Blinov2004}
B.~B. Blinov, D.~L. Moehring, L.~M. Duan, and C.~Monroe, Nature, \textbf{428},
  153 (2004).

\bibitem{Volz2006}
J.~Volz \emph{et~al.}, Phys. Rev. Lett., \textbf{96}, 030404 (2006).

\bibitem{Moehring2007}
D.~L. Moehring \emph{et~al.}, Nature, \textbf{449}, 68 (2007).

\bibitem{Togan2010}
E.~{Togan} \emph{et~al.}, Nature, \textbf{466}, 730 (2010).

\bibitem{Kimble2008}
H.~J. Kimble, Nature, \textbf{453}, 1023 (2008).

\bibitem{Ritter2012}
S.~Ritter \emph{et~al.}, Nature, \textbf{484}, 195 (2012).

\bibitem{Stute2012}
A.~Stute \emph{et~al.}, Nature, \textbf{485}, 482 (2012).

\bibitem{Wallentowitz1997}
S.~Wallentowitz, R.~L. de~Matos~Filho, and W.~Vogel, Phys. Rev. A, \textbf{56},
  1205 (1997).

\bibitem{Wallraff2005}
A.~Wallraff \emph{et~al.}, Physical Review Letters, \textbf{95}, 060501 (2005).

\bibitem{Mallet2009}
F.~Mallet \emph{et~al.}, Nat. Phys., \textbf{5}, 791 (2009).

\bibitem{Bianchetti2009}
R.~Bianchetti \emph{et~al.}, Physical Review A, \textbf{80}, 043840 (2009).

\bibitem{Walls1994}
D.~Walls and G.~Milburn, \emph{Quantum Optics} (Springer Verlag, Berlin, 1994).

\bibitem{Gardiner1985}
C.~W. Gardiner and M.~J. Collett, Phys. Rev. A, \textbf{31}, 3761 (1985).

\bibitem{Cahill1969}
K.~E. Cahill and R.~J. Glauber, Phys. Rev., \textbf{177}, 1857 (1969).

\end{thebibliography}
%
\end{document}